\documentclass{article}

\PassOptionsToPackage{numbers}{natbib}
\usepackage[preprint]{neurips_2026}
\usepackage{threeparttable}
\usepackage{amssymb}
\makeatletter
\renewcommand{\@bottomtitlebar}{
  \vskip 0.25in
  \vskip -\parskip
  \hrule height 1\p@
  \vskip 0.10in
  {\centering\large\scshape A Preprint\par}
  \vskip -0.10in          
}
\makeatother
\providecommand{\keywords}[1]{%
  \vspace{0.8em}\noindent\textbf{\textit{Keywords}}\hspace{0.6em}#1}
\usepackage{graphicx}

\usepackage[utf8]{inputenc} 
\usepackage[T1]{fontenc}    
\usepackage{hyperref}       
\usepackage{url}            
\usepackage{booktabs}       
\usepackage{amsfonts}       
\usepackage{nicefrac}       
\usepackage{microtype}      
\usepackage{xcolor}         

\title{MAPK Pathway Activity and Heme Biosynthesis Gene Expression in IDH-Wildtype Glioblastoma: A Purity-Adjusted, Discovery--Validation Analysis}

\author{%
  Manveer Singh Tib\\
  Independent Researcher\\
  \texttt{mnvertib@gmail.com} \\
}

\begin{document}

\maketitle
\vspace{-0.20in}          

\begin{abstract}
5-aminolevulinic acid (5-ALA) promotes fluorescence-based resection of glioblastoma via protoporphyrin IX (PpIX); however, there is considerable heterogeneity in fluorescence intensity, limiting margin distinction. Cell-line studies show that stimulation of the MAPK pathway results in decreased PpIX levels through increased elimination by ABCB1, an efflux transporter, and ferrochelatase (FECH), the enzyme that converts PpIX to heme. However, the mechanism has yet to be studied in human tissue. Using two independent, purity-adjusted sets of primary IDH-wildtype glioblastoma specimens (TCGA-GBM, n=140, discovery; CGGA, n=87, validation), no reproducible correlation between MAPK signaling and its proposed downstream effectors (ABCB1 and FECH) could be detected. Instead, MAPK activity displayed a reproducibly negative relationship with \textbf{PPOX}, the enzyme that converts protoporphyrinogen IX to protoporphyrin IX: TCGA-GBM (n=133 with purity estimates; $\rho = -0.40$, 95\% bootstrap CI $[-0.54, -0.25]$, adjusted $p < 0.001$), CGGA ($\rho = -0.28$, 95\% bootstrap CI $[-0.47, -0.07]$, adjusted $p = 0.0342$). The finding was robust to permutation testing and to alternative approaches to purity adjustment. Here we report, for the first time, replicated human-tissue evidence for this association. If this correlation reflects causality, MAPK regulation of PpIX would act at the point of synthesis rather than at the clearance steps implied by previous cell-culture studies. We observed differential activation of the MAPK pathway with increased activation in the classical subtype and decreased activation in the proneural subtype. There was no correlation between MAPK pathway activation and overall survival (log-rank $p = 0.73$). The association between PPOX expression and MAPK pathway activation is biological rather than prognostic, identifying the MAPK pathway as a candidate handle for increasing 5-ALA fluorescence.


\end{abstract}

\section{Introduction}

Glioblastoma is the most lethal and most common type of malignant brain tumor in adults; additionally, despite surgical, radiation and chemotherapy treatments \cite{stupp_radiotherapy_2005}, the prognosis remains poor. The most important prognostic factor that is within the surgeons' control is the extent of safe tumor resection. Therefore, researchers have developed intraoperative techniques to help distinguish neoplastic cells from normal tissue. One commonly employed technique is 5-aminolevulinic acid (5-ALA) guided surgery, which uses 5-ALA that when administered prior to surgery, is converted into protoporphyrin IX (PpIX), which selectively accumulates in glioblastoma tumors and emits red light under violet-blue illumination \cite{traylor_molecular_2021}. Randomized studies have shown that when PpIX can be visualized during surgery, this increases the amount of glioblastoma tumor that surgeons are able to remove and extends the time before the first signs of local recurrence or disease progression appear \cite{stummer_fluorescence-guided_2006}. This supports the use of 5-ALA fluorescence in addition to standard neurosurgical treatments for glioblastoma.

Protoporphyrin IX (PpIX) builds up in the body as a result of the heme biosynthesis pathway~\cite{traylor_molecular_2021, dailey_primer_2022}. Since 5-aminolevulinic acid (5-ALA), which is added from outside the body, enters into the heme biosynthesis pathway at a point after the rate limiting step, its conversion into protoporphyrinogen IX is not subject to the pathway's normal negative feedback mechanisms. PpIX is formed from protoporphyrinogen IX via oxidation by protoporphyrinogen oxidase (PPOX), while ferrochelatase (FECH) inserts an iron atom to form heme~\cite{dailey_primer_2022, obi_ferrochelatase_2022}. 

\begin{figure}[!t]
  \centering
  \includegraphics[width=0.72\columnwidth]{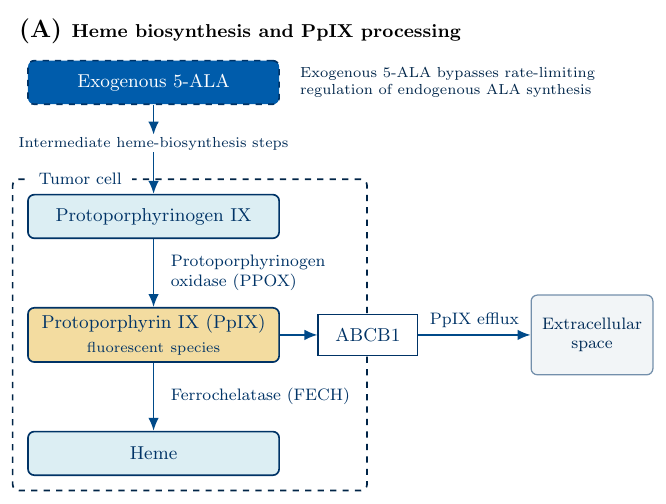}\\[0.7em]
  \includegraphics[width=0.72\columnwidth]{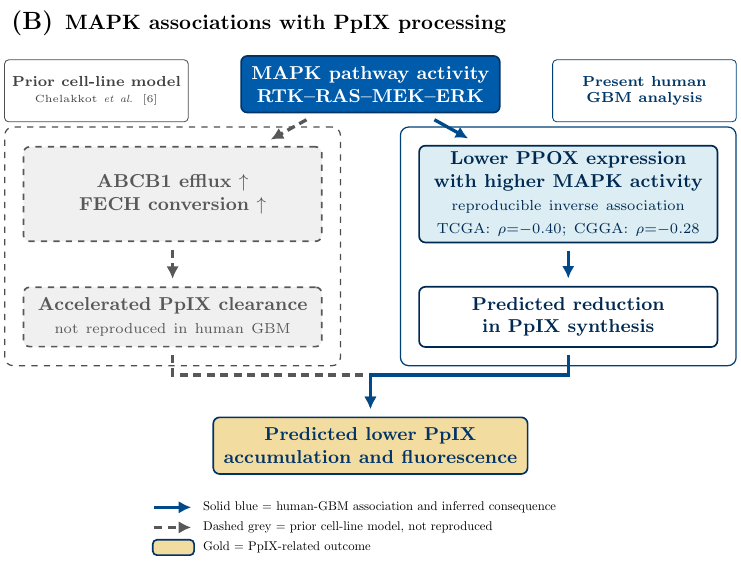}
  \caption{Heme biosynthesis and the hypothesized points of MAPK regulation.
  \textbf{(A)} Exogenous 5-ALA is metabolized downstream of the rate-limiting step;
  PPOX oxidizes protoporphyrinogen IX to fluorescent PpIX, after which FECH inserts
  iron to form heme. \textbf{(B)} In the \emph{in vitro} model of Chelakkot
  \emph{et al.}~\cite{chelakkot_mek_2020}, MAPK regulation acts on clearance
  (ABCB1 efflux; FECH conversion), whereas the present human-tissue data localize
  the reproducible association to synthesis (PPOX).}
  \label{fig:schematic}
\end{figure}

Thus, net PpIX accumulation reflects a steady-state equilibrium between its formation and subsequent processing and export, which are controlled by enzymes including PPOX and FECH and by porphyrin transporters including ABCB6, ABCG2, and the efflux pump ABCB1~\cite{chelakkot_mek_2020, kiening_recap_2022, schulz_abcb1_2023}. Clinically, this balance is not obtained: many tumors, and parts of individual tumors, do not contain enough fluorescent material, undermining the reliability of this method exactly where it is most needed~\cite{mischkulnig_heme_2022, almiron_bonnin_characterizing_2020}. Although the molecular basis of this variation has been investigated, it remains largely undefined~\cite{almiron_bonnin_characterizing_2020, pustogarov_hiding_2017}.

The potential ability of oncogenic signaling to influence how efficiently a tumor can produce fluorescent compounds from a supplied precursor compound such as 5-aminolevulinic acid (5-ALA)~\cite{yoshioka_enhancement_2018} must also be considered. Of all signaling axes affected in glioblastomas, the RTK/RAS/MEK/ERK (MAPK) pathway is the most frequently altered~\cite{the_cancer_genome_atlas_research_network_comprehensive_2008,yakubov_its_2025} and has been shown to control genes involved in processing PpIX by virtue of multiple in vitro and in vivo studies. MAPK signaling has been shown to upregulate two proteins involved in intraoperative fluorescence, the efflux transporter ABCB1 and FECH, which converts PpIX to heme~\cite{chelakkot_mek_2020, obi_ferrochelatase_2022}, thereby reducing the available PpIX to fluoresce. If this mechanism is applicable to humans, MAPK and its downstream effectors would be promising targets to increase fluorescence in faintly-fluorescing tumors.

The cell-culture findings may not be applicable to whole tumors, which consist of a mixture of cancerous and non-cancerous stromal and immune cells~\cite{zhang_tumor_2017,yoshihara_inferring_2013}. Therefore, to test the cell-line mechanism in intact tumor tissue, both tumor purity adjustment and independent replication cohorts are needed, as an association observed in only one cohort may be due to study-specific design or patient selection artifacts rather than true biology.

We used two independent, purity-adjusted datasets from separate patient populations, to test the MAPK--effector hypothesis \cite{chelakkot_mek_2020} in human glioblastoma. The originally proposed effector genes could not be reliably verified: there was no correlation between MAPK activity and ABCB1 expression in either dataset. In only one of the two datasets was a correlation between MAPK activity and FECH expression found. However, MAPK activity was consistently linked to reduced \textbf{PPOX} expression in both datasets. Therefore, the reproducible relationship lies one step upstream of the postulated effectors, at synthesis rather than clearance.

This study makes three principal contributions:
\begin{enumerate}
    \item First purity-adjusted two-cohort test of a previously preclinical MAPK--effector mechanism in human glioblastoma tissue.
    \item First replicated human-tissue evidence that MAPK activity is associated with the \emph{synthetic} step of the pathway (PPOX) and not the \emph{clearance} steps (ABCB1, FECH), repositioning the mechanistic locus proposed by cell-line studies.
    \item A clear empirical distinction between genes generalizable across cohorts (PPOX) and those not generalizable (ABCB1, FECH) that narrows the field of candidates for future functional and therapeutic investigation.
\end{enumerate}

\section{Related Work}
\subsection{Molecular Determinants of 5-ALA Fluorescence in Glioma}
5-ALA induced fluorescence is known to vary significantly within and between tumors and histological classification is a poor predictor of this heterogeneity~\cite{almiron_bonnin_characterizing_2020,nasir-moin_localization_2024}. Transcriptomic analyses have looked at the expression of genes in the heme pathway in fluorescent and non-fluorescent gliomas, sometimes with matched protein data~\cite{mischkulnig_heme_2022,mischkulnig_tcga_2020}, and have pointed to specific enzymes such as CPOX as possible regulators~\cite{pustogarov_hiding_2017}. Together, these studies suggest a relationship between heme biosynthetic gene expression and fluorescence, but are largely descriptive and do not test mechanistic hypotheses for the observed variation.

\subsection{Oncogenic Signaling as a Regulator of PpIX}
Ras/MEK/MAPK signaling decreases PpIX accumulation in cancer cells~\cite{yoshioka_enhancement_2018} by two distinct mechanisms discovered in a follow-up study~\cite{chelakkot_mek_2020}: MEK-mediated phosphorylation of RSK, which protects ABCB1 protein from proteasomal degradation, and MEK activation of HIF-1$\alpha$, which enhances FECH expression. The role of MAPK in the modulation of PpIX has been extensively studied in vitro and in vivo, but not in human glioblastoma tissue.

Here we quantified pathway activity using MPAS, a compact, well-validated transcriptional signature developed and clinically benchmarked across multiple cancer types~\cite{wagle_transcriptional_2018} and used in conjunction with ssGSEA enrichment of a curated hallmark gene set~\cite{barbie_systematic_2009, liberzon_molecular_2015, hanzelmann_gsva_2013} as an orthogonal cross-check (Section~\ref{sec:ppox}). There are other ways to infer pathway activity from transcriptomic data, including regulon- and transcription-factor-footprint-based methods. A systematic comparison of such methods against MPAS was beyond the scope of the present hypothesis-driven replication design, but is a natural direction for future work validating the generality of the PPOX association reported here.

\section{Dataset and Cohort Construction}

\subsection{Cohort Selection}
\label{sec:cohorts}
Two public glioblastoma RNA-seq datasets were analyzed independently of each other and expression matrices were never combined, allowing for independent discovery and validation. The Cancer Genome Atlas glioblastoma cohort (TCGA-GBM)~\cite{the_cancer_genome_atlas_research_network_comprehensive_2008} was used as the discovery dataset; and the Chinese Glioma Genome Atlas mRNAseq\_693 batch (CGGA)~\cite{zhao_chinese_2021} was used as the independent dataset to validate findings. Both cohorts were limited to IDH-wildtype primary CNS WHO grade 4 glioblastoma as defined by the 2021 WHO classification~\cite{zakharova_reclassification_2022}, which classifies IDH-mutant tumors as a separate entity, based on IDH/grade calls from Ceccarelli et al.~\cite{ceccarelli_molecular_2016} for TCGA and the clinical annotation provided for CGGA~\cite{zhao_chinese_2021}. By limiting both cohorts to one tumor grade and IDH genotype, we rule out grade-related variability in fluorescence intensity as a potential confound~\cite{traylor_molecular_2021}. Another quality control step removed 22 CGGA samples from further consideration due to abnormal library composition (i.e., RN7SL2 accounted for greater than 20\% of all reads). After removing samples from each dataset which were found to be non-qualifying, there were 140 samples remaining in the TCGA database and 87 samples remaining in the CGGA database (Table~\ref{tab:cohorts} for a summary of cohorts' demographics; exclusion criteria are described above).

\begin{table}[!ht]
  \caption{Characteristics of the discovery (TCGA-GBM) and validation (CGGA) cohorts, restricted to IDH-wildtype primary glioblastoma.}
  \label{tab:cohorts}
  \centering
  \begin{threeparttable}
    \begin{tabular*}{\columnwidth}{@{\extracolsep{\fill}}lcc}
    \toprule
    \textbf{Characteristic} & \textbf{TCGA} & \textbf{CGGA} \\
                            & (discovery)   & (validation)  \\
    \midrule
    N (IDH-WT primary GBM)& 140            & 87             \\
    Age, median [IQR]             & 62 [54--71]    & 56 [48--63]    \\
    Male, n (\%)                  & 89 (64\%)      & 49 (56\%)      \\
    Female, n (\%)\tnote{\ddag}   & 50 (36\%)      & 38 (44\%)      \\
    Follow-up months, median [IQR]& 9 [5--15]      & 14 [8--25]     \\
    Deaths, n (\%)                & 94 (67\%)      & 74 (85\%)      \\
    MGMT methylated, n (\%)\tnote{*} & --          & 45 (52\%)      \\
    ABCB1 detection\tnote{\dag}   & 100\% (8.40)   & 100\% (3.06)   \\
    FECH detection\tnote{\dag}    & 100\% (9.40)   & 100\% (3.11)   \\
    \bottomrule
    \end{tabular*}
    \begin{tablenotes}
      \item[*] MGMT promoter methylation status is not part of the curated pan-glioma annotation source used in this study~\cite{ceccarelli_molecular_2016}; TCGA-GBM MGMT status has been reported elsewhere using methylation array-derived calls not included in the present analysis.
      \item[\dag] Detection is the percentage of samples that have gene expression values above zero. The number in parentheses indicates the median expression on the native scale of each dataset and is not comparable across datasets.
      \item[\ddag] Sex was not available for one TCGA sample; percentages are of the 139 samples with recorded sex.
    \end{tablenotes}
  \end{threeparttable}
\end{table}

\subsection{Data Acquisition and Preprocessing}
The expression levels in TCGA were $\log_2$-normalized at the gene level using HiSeqV2 and clinical data were retrieved from UCSC Xena~\cite{goldman_visualizing_2020}. Estimates of tumor purity derived from ABSOLUTE along with molecular subtyping of gliomas (molecular subtypes) were obtained from a pan-glioma resource~\cite{ceccarelli_molecular_2016}. The CGGA has also made available both STAR/RSEM-based gene-expression levels and associated clinical data~\cite{zhao_chinese_2021}, $\log_2$ transformed with +1 added to each value ($\log_2(x+1)$). Because the two cohorts' expression matrices were acquired, IDH-filtered, and normalized independently of one another (Section~\ref{sec:cohorts}), the possibility of information leakage between the datasets was removed.

\section{Methodology}
\subsection{MAPK Pathway Activity Quantification}
\label{sec:mpas}
The level of activation of the MAPK/ERK pathway was determined for each sample with the MAPK Pathway Activity Score (MPAS)~\cite{wagle_transcriptional_2018}, based upon the average z-score of the ten known transcriptional targets of the MAPK/ERK pathway (DUSP4, DUSP6, ETV4, ETV5, PHLDA1, SPRY2, SPRY4, CCND1, EPHA2, EPHA4). Since z-scores are normalized to each dataset, separate MPAS scores were generated for each cohort to prevent cross-platform scale differences from being introduced into the comparison of cohorts. To validate this measure of pathway activity, an additional index was created using single-sample Gene Set Enrichment Analysis (ssGSEA)~\cite{barbie_systematic_2009} of the \texttt{HALLMARK\_KRAS\_SIGNALING\_UP} gene set~\cite{liberzon_molecular_2015}, which is part of the Molecular Signatures Database (MSigDB). ssGSEA was computed through the GSVA package (version 2.6.2)~\cite{hanzelmann_gsva_2013}.

\subsection{Effector and Heme-Pathway Genes}
Expression levels for already identified downstream target genes (ABCB1 and FECH)~\cite{chelakkot_mek_2020} as well as others in the heme pathway (ALAD, HMBS, UROS, UROD, CPOX, \textbf{PPOX}, ABCB6, ABCG2, SLC15A1 and SLC15A2)~\cite{mischkulnig_tcga_2020, dailey_primer_2022} were assessed in all sample sets. SLC15A1 was evaluated using only TCGA data due to the absence of this gene from the CGGA expression matrix. We first confirmed that both ABCB1 and FECH were detectable in 100\% of samples in each dataset prior to evaluation.

\subsection{Tumor-Purity Adjustment}
We normalized all primary correlations for tumor purity, as bulk transcriptomes contain signal from both cancerous cells and the microenvironment~\cite{zhang_tumor_2017}. Purity estimation was performed by ABSOLUTE (133 out of 140 samples with available values)~\cite{ceccarelli_molecular_2016}, the remaining seven samples were discarded. Since CGGA does not have the somatic copy-number data required by ABSOLUTE, we estimated purity using ESTIMATE v1.0.13~\cite{yoshihara_inferring_2013}, a composite of its stromal and immune components. Adjustments for each cohort's purity adjusted association estimates were performed via partial Spearman correlation utilizing the ppcor R-package (version 1.1)~\cite{kim_ppcor_2015}, where each cohort's specific purity estimate was employed as the controlling variable.

\subsection{Statistical Analysis}
Correlations between MPAS and gene expression were calculated with Spearman correlation and purity-adjusted partial Spearman correlation. Multiple hypothesis testing correction was performed on all tested genes in the panel by applying the Benjamini-Hochberg adjustment~\cite{benjamini_controlling_1995}. To determine if there are differences in MPAS values for the four GBM transcriptional subtypes (Proneural, Classical, Mesenchymal, Neural)~\cite{verhaak_integrated_2010, wang_tumor_2017} identified in the TCGA dataset, we used the Kruskal-Wallis test. The Unassigned (n=12) subtypes were removed from this analysis as a result of the lack of subtyping. Due to the lack of subtype information about the CGGA dataset, we could therefore only focus on the subtypes found in the TCGA. The overall survival of TCGA patients was estimated using the Kaplan-Meier method. We separated the MPAS values at the median and grouped them into MEK-high and MEK-low and then used a log-rank test to see if there were any statistical differences. We also created a Cox proportional hazard model that had MPAS as the primary independent variable with age and sex as secondary independent variables to evaluate survival. Confidence intervals for the gene-panel correlations shown in Figure~\ref{fig:hemepanel} were obtained by Fisher $z$-transformation of the partial Spearman $\rho$, using standard error $1/\sqrt{n-k-3}$ with $k=1$ controlling variable, back-transformed to the correlation scale. As an additional robustness check, 10,000-resample percentile bootstrap confidence intervals were computed separately for PPOX and are reported in Section~\ref{sec:robustness}.

\subsubsection{Robustness analyses}
Sensitivity analyses using four approaches were conducted for both cohorts to assess the
inverse relationship between the MPAS and expression levels of the \textbf{PPOX} gene. These
analyses used the same purity-adjusted partial Spearman statistic that was used in the
primary analysis. 95\% CIs were derived using case-resampling bootstrap (resampling individuals with replacement, $R = 10,000$), using the 2.5th and 97.5th percentiles of the resulting distribution. We generated a label-permutation null by permuting the PPOX expression 10,000 times with MPAS and the purity covariate held fixed, recomputing the partial Spearman $\rho$ for each permutation. The two-sided empirical p-value was $(\#\{|\rho^{*}| \geq |\rho_{\mathrm{obs}}|\} + 1)/(n_{\mathrm{perm}} + 1)$. The effect of different purity-correction approaches on the relationship was assessed through comparison of covariates: TCGA-GBM had uncorrected, ABSOLUTE, and ESTIMATE-combined; CGGA had uncorrected, ESTIMATE combined, stromal-only, and immune-only. Finally, MPAS was recalculated ten times per cohort, each time omitting one of its ten constituent genes, to test whether the association depended on any single gene in the score. The primary 12-gene panel test (Table~\ref{tab:table2}) is subject to the Benjamini--Hochberg correction as described above within each cohort. Additional robustness analyses (bootstrap resampling, label permutation, alternative purity-adjustment covariates, and leave-one-gene-out re-runs of MPAS) are sensitivity checks on the single PPOX association already identified as significant, and are reported without additional correction, as is standard for confirmatory robustness testing of a pre-specified primary finding. As an exploratory check on whether the transcript-level PPOX association also holds at the protein level, an independent proteogenomic cohort (CPTAC-GBM, $n=99$) with matched RNA-seq and mass-spectrometry proteome data was additionally examined (results reported in Section~\ref{sec:limitations}). All data analyses were performed using R.

\section{Results}
\subsection{Cohort Characteristics}
We evaluated 140 TCGA and 87 CGGA IDH wildtype primary GBM samples that passed filtering. The cohorts were demographically similar (Table~\ref{tab:cohorts}). ABCB1 and FECH mRNA were detected in 100\% of samples in both cohorts.

\begin{table}[!ht]
  \caption{Purity-adjusted partial Spearman correlations between MAPK activation (MPAS) and heme biosynthesis gene expression in discovery (TCGA) and validation (CGGA) datasets.}
  \label{tab:table2}
  \centering
  \footnotesize
  \begin{threeparttable}
    \renewcommand{\arraystretch}{1}
    \begin{tabular*}{\columnwidth}{@{\extracolsep{\fill}}lccccc}
    \toprule
    & \multicolumn{2}{c}{\textbf{TCGA (discovery)}} & \multicolumn{2}{c}{\textbf{CGGA (validation)}} & \\
    \cmidrule(lr){2-3} \cmidrule(lr){4-5}
    \textbf{Gene} & $\rho$ & BH $q$ & $\rho$ & BH $q$ & \textbf{Repl.}\tnote{\ddag} \\
    \midrule
    ABCB1   & $0.07$  & $0.540$   & $0.13$  & $0.331$ & \\
    FECH    & $0.08$  & $0.515$   & $0.28$  & $0.0342$ & \\
    ALAD    & $-0.40$ & $<0.001$  & $-0.14$ & $0.331$ & \\
    HMBS    & $-0.19$ & $0.071$   & $-0.06$ & $0.611$ & \\
    UROS    & $-0.16$ & $0.113$   & $-0.38$ & $0.003$ & \\
    UROD    & $-0.07$ & $0.540$   & $-0.25$ & $0.0498$ & \\
    CPOX    & $0.17$  & $0.102$   & $0.09$  & $0.473$ & \\
    PPOX    & $-0.40$ & $<0.001$\tnote{\dag}  & $-0.28$ & $0.0342$\tnote{\dag} & $\checkmark$ \\
    ABCB6   & $-0.24$ & $0.023$   & $-0.22$ & $0.083$ & \\
    ABCG2   & $-0.02$ & $0.868$   & $0.25$  & $0.0498$ & \\
    SLC15A1 & $0.05$  & $0.635$   & --      & --      & \\
    SLC15A2 & $-0.21$ & $0.0455$  & $-0.06$ & $0.611$ & \\
    \bottomrule
    \end{tabular*}
    \begin{tablenotes}
      \footnotesize
      \item Values are purity-adjusted partial Spearman correlation coefficients ($\rho$) between MPAS and gene expression, with Benjamini--Hochberg (BH) adjusted $q$-values, computed across the gene panel within each cohort (TCGA: ABSOLUTE purity, $n=133$; CGGA: ESTIMATE score, $n=87$). Only \textbf{PPOX} was inversely correlated with MPAS in both datasets; SLC15A1 was not found in CGGA.
      \item[\ddag] Replicated: BH $q<0.05$ in both cohorts (pre-specified criterion).
      \item[\dag] Raw (pre-BH) $p$-values for PPOX, fed into Fisher's combined test (Section~\ref{sec:ppox}): TCGA $p=1.72\times 10^{-6}$; CGGA $p=9.32\times 10^{-3}$.
      \item BH-adjusted $q$ values are reported to three significant figures. ABCG2 and UROD have the same adjusted value (0.0498) because they are adjacent in the ranked $p$-value list and the Benjamini--Hochberg step-up procedure assigns both the same cumulative minimum.
    \end{tablenotes}
  \end{threeparttable}
\end{table}

\subsection{The Pre-Specified Effector Associations Were Not Supported}
\begin{figure}[!ht]
  \centering
  \includegraphics[width=0.90\columnwidth]{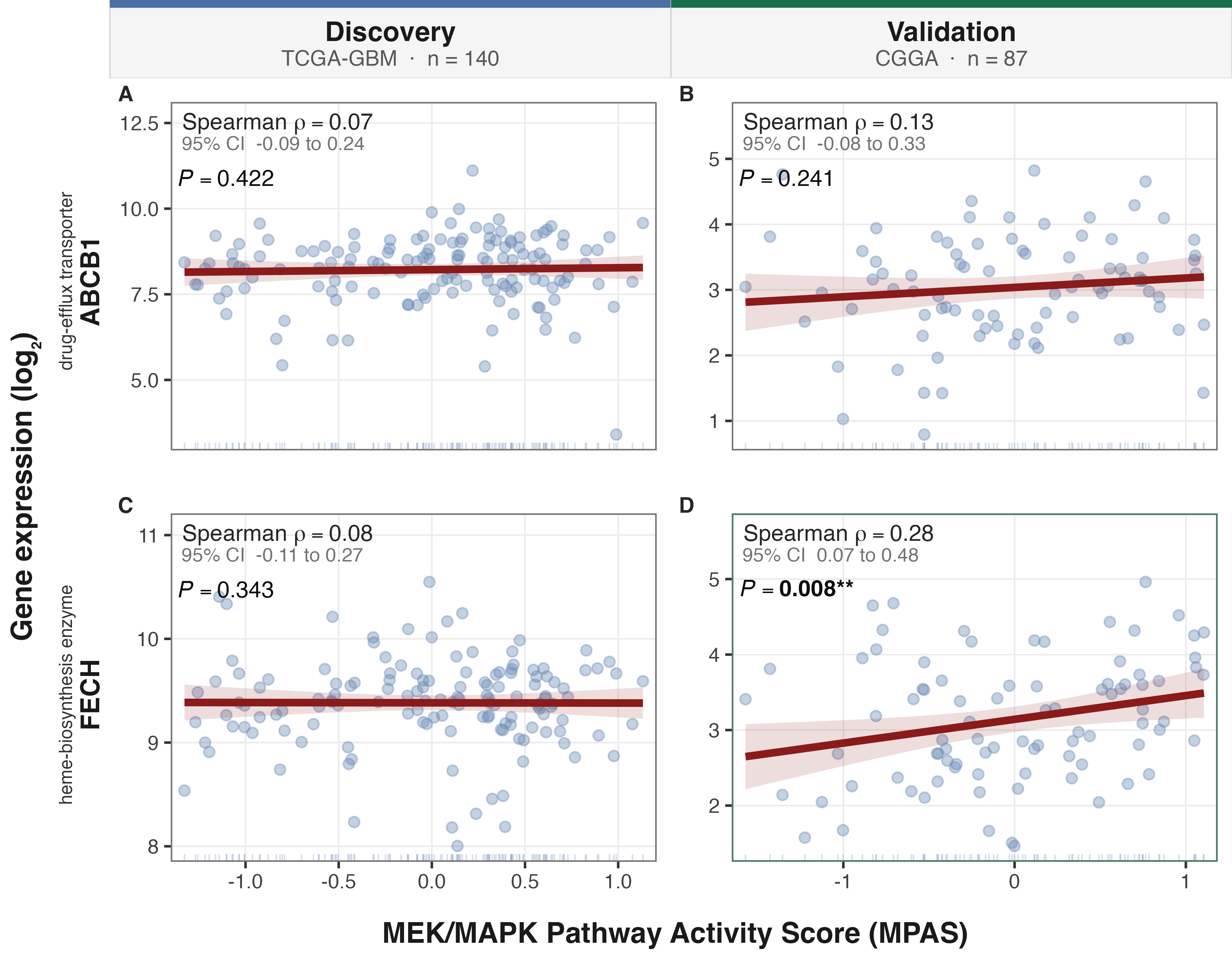}
    \caption{Purity-adjusted partial Spearman associations between MPAS and the predicted effectors ABCB1 and FECH (TCGA discovery; CGGA validation). Points are samples; lines are linear fits with 95\% CIs. $P$ values shown in the panels are unadjusted; Benjamini--Hochberg adjusted $q$-values for these genes are given in Table~\ref{tab:table2}.}
  \label{fig:effectors}
\end{figure}

The predicted relationship between high MAPK activity and increased expression of the PpIX-processing genes (ABCB1 and FECH) that was demonstrated through the mechanistic cell line model \cite{chelakkot_mek_2020} was not observed in this research. Following purity adjustments, there was no evidence of an association for either effector with MPAS in TCGA ($\rho = 0.07$; BH $q = 0.540$; $\rho = 0.08$; BH $q = 0.515$ for ABCB1 and FECH respectively). A similar lack of association occurred for ABCB1 in CGGA ($\rho = 0.13$; BH $q = 0.331$), although an association between MPAS and FECH was noted ($\rho$ = 0.28; BH-adjusted $q = 0.0342$). Therefore the previously established relationship was not seen in the discovery cohort and the only gene-level effector association -- a positive association for FECH -- only occurred in the validation cohort (Fig.~\ref{fig:effectors}).

To test for consistency of the findings, the relationship between the KRAS signaling ssGSEA score and FECH was also examined as an independent measure. The correlation was again positive for ssGSEA KRAS-signaling in CGGA ($\rho$ = 0.43; p = $4.1 \times 10^{-5}$), but no correlation was found in TCGA ($\rho$ = -0.06; p = 0.51). Thus, while there is a consistent relationship between FECH and ssGSEA KRAS-signaling in the CGGA cohort, the relationship was not generalizable across both datasets.

\subsection{MAPK Activity Is Reproducibly Associated with Lower PPOX Expression}
\label{sec:ppox}
\begin{figure}[!ht]
  \centering
  \includegraphics[width=0.92\columnwidth]{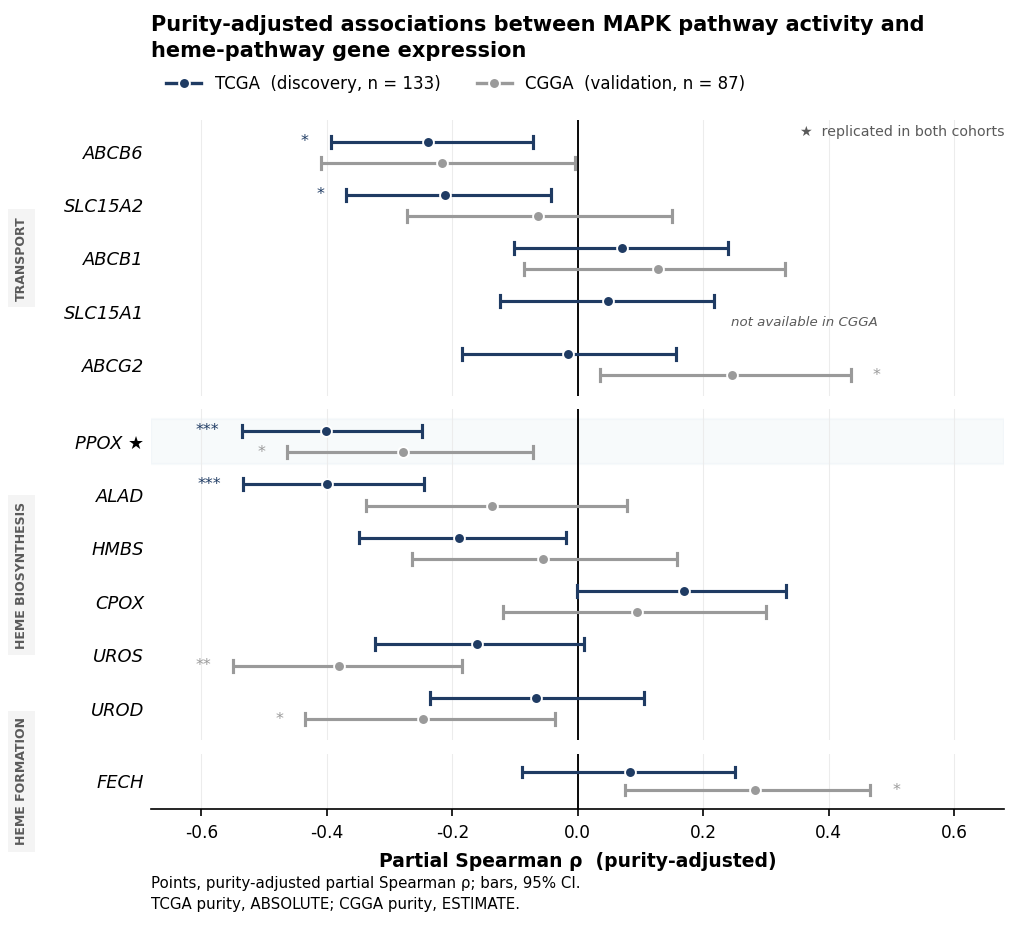}
  \caption{\textbf{Purity-adjusted associations between MAPK pathway activity and
  heme-pathway gene expression.} Points denote the purity-adjusted partial Spearman
  $\rho$ between MAPK pathway activity score (MPAS) and gene expression; horizontal
  bars denote 95\% confidence intervals obtained by Fisher $z$-transformation.
  Partial Spearman correlations were computed with tumor purity as the controlling
  variable. Tumor purity was estimated by ABSOLUTE in TCGA and by ESTIMATE in CGGA.
  $P$ values were adjusted by the Benjamini--Hochberg procedure within each cohort,
  across 12 genes in TCGA and the 11 genes available in CGGA.
  \textbf{*, BH-adjusted $P<0.05$; **, BH-adjusted $P<0.01$; ***, BH-adjusted
  $P<0.001$.} $\star$ denotes replication in both cohorts, defined as BH-adjusted
  $P<0.05$ with concordant direction of effect; PPOX was the only gene meeting this
  criterion. SLC15A1 was absent from the CGGA expression matrix and was tested in
  TCGA only. Genes are grouped by function (transport, heme biosynthesis, heme
  formation) and ordered within group by the magnitude of the TCGA correlation.
  TCGA $n=133$ (samples with available ABSOLUTE purity estimates); CGGA $n=87$.}
  \label{fig:hemepanel}
\end{figure}

The only one of the heme biosynthetic genes that had a statistically significant inverse correlation with MPAS in both groups was \textbf{PPOX} (protoporphyrinogen oxidase) which is responsible for the penultimate reaction of the pathway (Fig.~\ref{fig:hemepanel}). The inverse association was also significant after correction for tumor purity and Benjamini-Hochberg correction in both TCGA ($\rho$ = -0.40; BH $q < 0.001$) and CGGA ($\rho$ = -0.28; BH $q = 0.0342$). Fisher's combination of the purity-adjusted, pre-multiplicity partial-correlation $p$-values of the two cohorts is $\chi^2 = 35.90$ (df $= 4$), $p = 3.0 \times 10^{-7}$, showing that the joint evidence for the PPOX association is substantially stronger than either cohort alone and is unlikely to be a chance survivor of multiple testing. Six further heme-pathway genes, excluding the two pre-specified effectors, reached BH-adjusted $q < 0.05$ in one cohort but not the other: ALAD, ABCB6, and SLC15A2 in TCGA, and UROS, UROD, and ABCG2 in CGGA (Table~\ref{tab:table2}). As such, we found PPOX to be the only gene from the heme biosynthetic pathway that demonstrated an inverse relationship with MPAS reproducible across the two independent glioma study groups, and --- to our knowledge --- the first gene-level association between MAPK signaling and heme biosynthesis activity to reproduce across independent glioma cohorts.

To independently confirm that this association is not an artifact of the particular ten-gene composition of MPAS, we repeated the analysis using ssGSEA enrichment of the \texttt{HALLMARK\_KRAS\_SIGNALING\_UP} gene set (Section~\ref{sec:mpas}) as an alternative measure of pathway activity. Using this orthogonal metric, we replicated the inverse association with PPOX in both cohorts (TCGA: $\rho = -0.29$, 95\% CI $[-0.44, -0.13]$, $p = 7.7 \times 10^{-4}$; CGGA: $\rho = -0.65$, 95\% CI $[-0.76, -0.51]$, $p = 1.3 \times 10^{-11}$). The CGGA result was robust to exclusion of the most extreme observations ($\rho = -0.62$ with the four most extreme points dropped). The larger effect size in CGGA under this alternative metric than MPAS ($\rho = -0.28$) suggests some sensitivity to the specific instrument used to measure pathway activity, although the direction and significance of the association were consistent across both metrics and both cohorts.

\subsection{Robustness of the PPOX association}
\label{sec:robustness}
As a test for the stability of the discovered inverse correlation between MPAS and \textbf{\textit{PPOX}}, and to rule out any potential effect of multiplicity or analysis choice, four robustness analyses were performed for each dataset.

The first consisted of a case-resampling bootstrap with 10,000 resamples, resulting in a
$95\%$ confidence interval for the purity-adjusted partial Spearman correlation that did
not include zero in either dataset: TCGA-GBM $\rho = -0.402$ (95\% CI $[-0.542, -0.247]$);
CGGA $\rho = -0.279$ (95\% CI $[-0.471, -0.071]$).

Moreover, a label-permutation test with 10,000 iterations showed that the measured
correlations were very rare events in the tail of their null distributions. The observed
$\rho$ in TCGA-GBM was $-4.6$ standard deviations away from the mean of its null
distribution, with no label permutation showing a result as extreme as the observed
correlation (two-tailed empirical $p < 1\times10^{-4}$). In CGGA, the observed $\rho$ was
$-2.5$ standard deviations away from the mean of its null distribution ($p = 0.010$).

Lastly, the MPAS--\textit{PPOX} relationship was consistently negative and statistically
significant in all scenarios of purity adjustment --- unadjusted, ABSOLUTE, and ESTIMATE
combined for TCGA-GBM, and unadjusted, ESTIMATE combined, stromal-only, and immune-only for CGGA --- across all seven cohort-by-method combinations (all $p < 0.05$ after Benjamini--Hochberg correction). Importantly, when both cohorts are controlled with the same ESTIMATE-combined purity metric, eliminating the possibility that the cross-cohort replication is due to a difference in purity estimation method rather than biology, the inverse MPAS--\textit{PPOX} association is still negative and statistically significant in both TCGA-GBM and CGGA. In addition, MPAS was recalculated ten times for each of the cohorts; each time, a different gene from its ten-gene set was excluded. The relationship with PPOX remained negatively correlated and statistically significant, regardless of which gene was omitted, for both TCGA ($\rho$ range: $-0.45$ to $-0.36$, all $p < 0.001$) and CGGA ($\rho$ range: $-0.33$ to $-0.25$, all $p < 0.05$).

The inverse MPAS--\textbf{\textit{PPOX}} correlation also showed directional similarity in an independent CPTAC-GBM proteomic cohort ($n=99$; $\rho = -0.13$, 95\% CI $[-0.32, 0.07]$, $p = 0.21$), though it did not reach statistical significance; see Section~\ref{sec:limitations} for discussion.

Collectively, these analyses suggest that the MPAS--\textbf{\textit{PPOX}} relationship is stable
and reliable against a variety of analyses. As mentioned elsewhere, this does not change the fact that the relationship is a transcript-level relationship of modest effect size. A large effect seen in one dataset, where cohort-specific technical or population artifacts cannot be excluded, is weaker evidence for a true biological signal than a small effect size replicated across two independent, differently sourced and separately normalized cohorts.

\subsection{MAPK Activity Distinguishes Transcriptional Subtypes}
\begin{figure}[!ht]
  \centering
  \includegraphics[width=0.92\columnwidth]{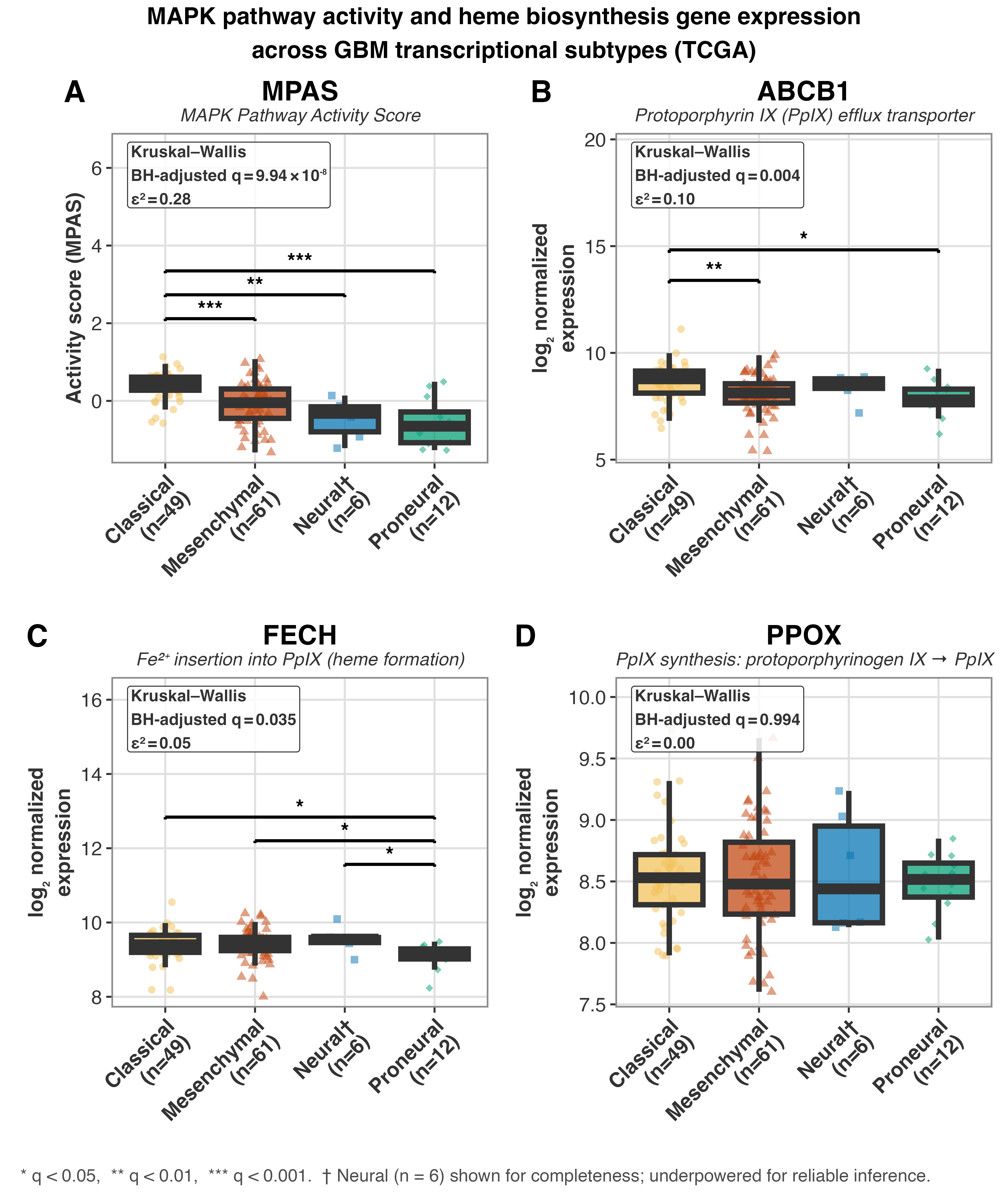}
\caption{\textbf{Heme biosynthesis gene expression and MAPK pathway activity
  across GBM transcriptional subtypes (TCGA).} \textbf{(A)} MPAS, \textbf{(B)}
  ABCB1, \textbf{(C)} FECH, and \textbf{(D)} PPOX by Verhaak
  subtype~\cite{verhaak_integrated_2010}. Boxes indicate median and interquartile
  range; individual samples are overlaid points. The omnibus tests are
  Kruskal--Wallis with Benjamini--Hochberg correction across the four panels,
  reported as BH-adjusted $q$; $\varepsilon^2$ is the corresponding effect size.
  Pairwise comparisons are Dunn's test with Benjamini--Hochberg correction, and
  only comparisons reaching $q < 0.05$ are shown. The Neural subtype ($n=6$) is
  included for completeness but is underpowered for reliable inference. PPOX
  exhibits no detectable difference in expression at the subtype level (D),
  indicating that the continuous MAPK--PPOX association reported in
  Fig.~\ref{fig:hemepanel} is not explained by subtype-level differences.}
  \label{fig:subtypes}
\end{figure}

MPAS differed significantly among the four major transcriptional subtypes defined by Verhaak et al.~\cite{verhaak_integrated_2010} in the TCGA dataset (Kruskal--Wallis BH-adjusted $q = 9.94 \times 10^{-8}$, $\varepsilon^2 = 0.28$; Fig.~\ref{fig:subtypes}). The highest MPAS was found in the classical subtype and the lowest in the proneural subtype. ABCB1 ($q = 0.004$, $\varepsilon^2 = 0.10$) and FECH ($q = 0.035$, $\varepsilon^2 = 0.05$) were also found at their highest mRNA levels in the classical subtype, following the same pattern. In contrast, there was no statistically significant difference in \textbf{PPOX} expression between subtypes ($q = 0.994$, $\varepsilon^2 = 0.00$).

\subsection{MAPK Activity Is Not Prognostic for Overall Survival}
\begin{figure}[!ht]
  \centering
  \includegraphics[width=0.90\columnwidth]{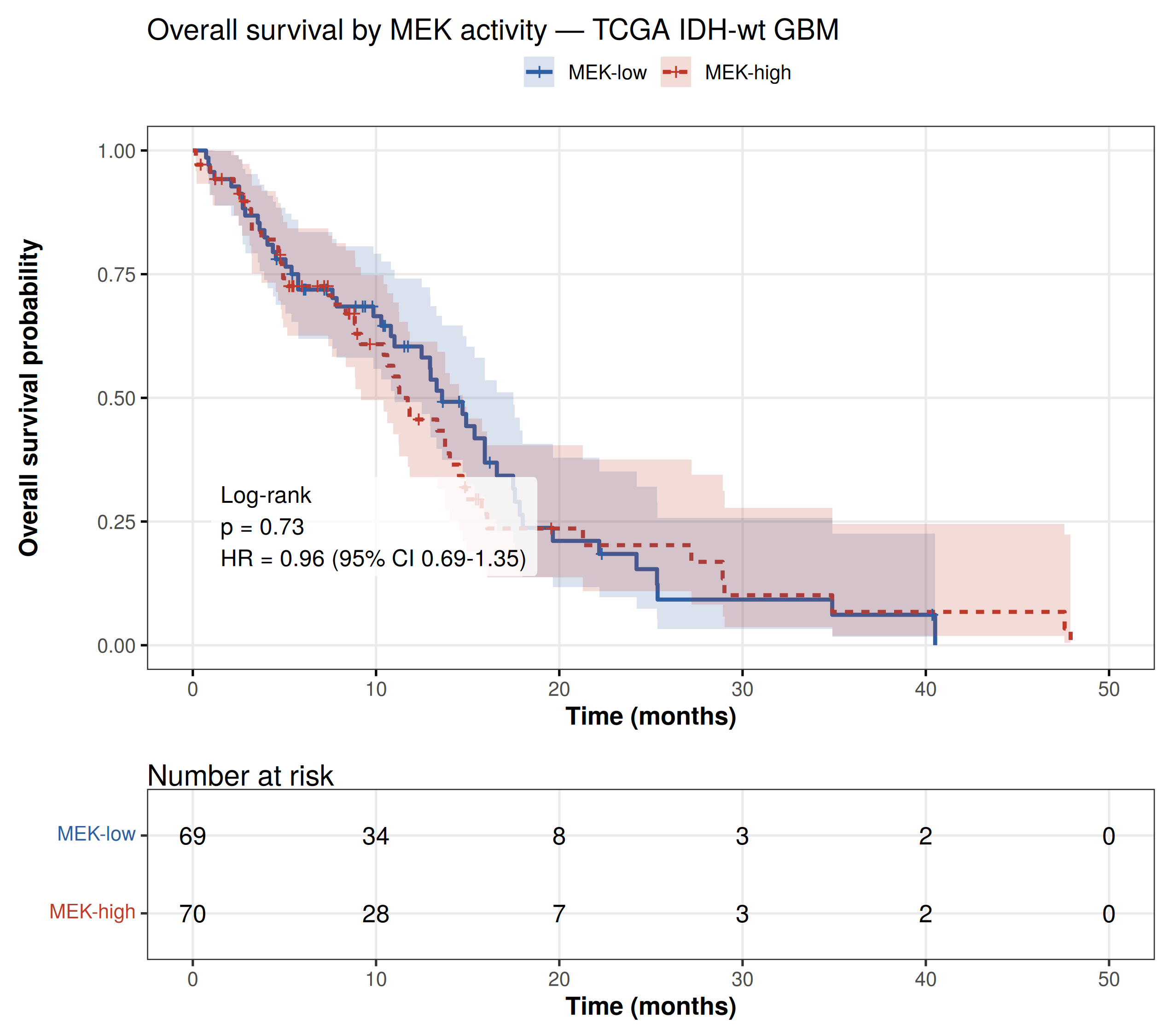}
  \caption{MAPK pathway activity is not linked to overall survival in TCGA IDH-wildtype glioblastoma. Kaplan--Meier curves for MEK-low ($n=69$) and MEK-high ($n=70$) patients, split by median MPAS value (log-rank $p=0.73$). Censoring is indicated by tick marks. Numbers at risk are given below.}
  \label{fig:survival}
\end{figure}

The median MPAS values from TCGA data did not result in a difference in Overall Survival (log-rank p = 0.73; Fig.~\ref{fig:survival}) when the patient population was split at this median MPAS value to create two groups of patients, MEK-low (n=69) and MEK-high (n=70). Multivariable Cox regression revealed that MPAS was not predictive of survival (HR = 0.96 per unit MPAS; 95\% CI 0.69--1.35; p = 0.83). Age was found to be statistically significantly predictive of survival (HR = 1.03 per year; 95\% CI 1.01--1.05; p = 0.011), while sex was not predictive (p = 0.66). Using Schoenfeld residuals, we tested the proportional hazards assumption and found no evidence that MPAS, age, or sex violated the proportionality assumption (global $\chi^2 = 1.65$, $df = 3$, $p = 0.65$).

\section{Discussion}
\subsection{A Reproducible Signal Relocated from Clearance to Synthesis}
In the current study, we provide for the first time replicated human-tissue evidence that localizes the consistent MAPK--heme-pathway association at the synthetic step (PPOX) rather than at clearance. MAPK activity was reproducibly linked to lower PPOX expression, the only heme biosynthesis gene that met the pre-specified replication criterion, in both cohorts. The original hypothesized effectors of this relationship, ABCB1 and FECH \cite{chelakkot_mek_2020}, were not confirmed to the same standard. No association between MAPK activity and ABCB1 was found in either cohort, a gap consistent with RSK's reported role in protecting extant ABCB1 protein from proteasomal degradation rather than inducing its transcription \cite{chelakkot_mek_2020}, a post-translational mechanism that this study's transcript-level design cannot detect. There was an association between MAPK activity and FECH, but it was only in one of the two cohorts and does not meet the same replication criterion as PPOX.

The mechanism itself is not invalidated by this finding; rather, the reproducible signal sits further upstream in the pathway, at PpIX synthesis. PPOX synthesizes PpIX from protoporphyrinogen IX, so if there is any causality in the relationship, reduced PPOX in tumors with greater MEK/MAPK activation would lower PpIX synthesis rather than accelerate its clearance. Both pathways point in the same direction -- less PpIX when MAPK signaling is high -- but this direction is now maintained by a different, reproducible node than initially suggested. Previous descriptive heme-pathway studies in glioma have not used a pre-specified replication criterion in a symmetric fashion across independent cohorts.

The position of PPOX may explain why this signal can be detected only for this one of the six heme-biosynthesis genes tested. It catalyzes the terminal oxidation of protoporphyrinogen IX to PpIX, the immediate biosynthetic step before the fluorescent molecule, without any intermediate conversions. It should be pointed out that PPOX did not differ in terms of the size of the correlation coefficient: ALAD exhibited an equally strong negative correlation in TCGA ($\rho = -0.40$), and several other heme-pathway genes trended negative in one of the two datasets. What distinguishes PPOX is that it reached significance in both cohorts, independently obtained and separately normalized, which was precisely the criterion we established beforehand. This pattern nonetheless remains compatible with MPAS being broadly anti-correlated with metabolic and biosynthetic transcription in these tumors; a genome-wide comparison of MPAS--gene correlations, or adjustment for a proliferation signature, would be needed to establish that the PPOX association is specific rather than one instance of a general trend. We note that this mechanistic account depends on transcript-level associations. The analogous protein-level test (Section~\ref{sec:limitations}) showed the same direction of effect, but did not reach significance. The synthesis-based explanation should thus be taken as the best-supported hypothesis, not an established mechanism. A decrease in flux at this step means that less of the fluorescent species is made. In contrast, a similar decrease earlier in the pathway (e.g., at ALAD, UROS or CPOX) could in principle be buffered by accumulation and processing of upstream intermediates before PpIX. This argument is compatible with a rate controlling role for PPOX but does not prove it. In general, the MAPK/ERK pathway regulates expression of many metabolic and biosynthetic genes, mainly through ETS-family and AP-1 transcription factors. Whether these factors directly regulate the transcription of PPOX still needs to be demonstrated by future chromatin or reporter assays. 

The consequences of PPOX reduction are also not straightforward in direction. Pharmacological PPOX inhibition leads to accumulation of protoporphyrinogen IX, which leaks from the mitochondria into the cytosol and auto-oxidizes to form PpIX, thus increasing PpIX formation despite the enzyme blockade~\cite{dailey_primer_2022}. Whether transcriptional reduction of PPOX produces the same result remains open, given that the leakage mechanism requires substrate accumulation exceeding mitochondrial retention, which may not be reached under transcriptional downregulation. If direct transcriptional regulation of PPOX by MAPK is confirmed, this would change the therapeutic logic of the pathway: rather than a peripheral transporter (ABCB1) or a downstream converting enzyme (FECH), the handle posited by this analysis is at the ultimate synthetic step of the fluorescent product itself -- in line with our suggestion that MEK/MAPK inhibition, rather than direct augmentation of PPOX, is the more tractable translational target (Section~\ref{sec:conclusion}).

In order to distinguish between decreased synthesis and increased clearance, a direct measurement of PpIX levels would be needed, which is not possible within the gene-expression-only design.

\subsection{Discordance at FECH}
The FECH result needs to be validated independently, since a significant MAPK--FECH correlation was found in CGGA (using the KRAS signaling score) but not in TCGA. Possible reasons for the discrepancy include differences in sequencing platform, normalization, patient ancestry~\cite{zhao_chinese_2021} or tumor microenvironment. No single factor explains the discrepancy, and therefore no generalization regarding FECH is justified. With our pre-specified criterion requiring replication in both datasets, the MAPK--FECH association is not robust while the MAPK--\textbf{PPOX} association is.

\subsection{Subtype Independence and Absence of Prognostic Value}
There was significant variation in MAPK activity among the four subtypes, with the highest in classical and the lowest in proneural. However, levels of \textbf{PPOX} mRNA appear to be almost equivalent in all four subtypes. Thus, the association seen between PPOX and MAPK could not reasonably be attributed to differential gene expression that distinguishes the subtypes. To infer whether the subtype-level association of MAPK activity with ABCB1/FECH expression corresponds to a true continuous association obscured by the subtype groupings, we also examined the MPAS--ABCB1 and MPAS--FECH correlations within each individual transcriptional subtype and further re-implemented the main partial correlation with subtype as a covariate in addition to tumor purity. Within-subtype correlations were weak and non-significant for both genes in the two largest subtypes (ABCB1 -- Classical: $\rho=-0.03$, $p=0.84$; Mesenchymal: $\rho=0.03$, $p=0.80$; FECH -- Classical: $\rho=0.11$, $p=0.46$; Mesenchymal: $\rho=0.10$, $p=0.44$). The Neural ($n=6$) and Proneural ($n=12$) subtypes were underpowered for reliable within-group inference. Adding subtype as an additional covariate with purity did not recover an association for either gene (ABCB1: $\rho = -0.03$, $p = 0.71$; FECH: $\rho = 0.05$, $p = 0.60$; $n=121$). This suggests that the subtype-specific co-elevation of MAPK activity with ABCB1 and FECH reflects two distinct subtype-specific readouts, rather than a continuum, in line with these genes being subtype/lineage markers rather than direct MAPK targets. In addition, no relationship between patient survival and MAPK activity has been demonstrated using log-rank or multivariable Cox regression analysis. Only increased age was associated with poorer survival in the latter model. Therefore, the PPOX--MAPK relationship likely reflects the transcriptional control of PpIX biosynthesis and does not appear to be a useful indicator of prognosis.

\subsection{Limitations and Future Directions}
\label{sec:limitations}
Bulk RNA sequencing measures signal from malignant and stromal cells, and purity adjustment reduces but cannot eliminate this confounding~\cite{yoshihara_inferring_2013}. Transcript abundance does not account for post-translational modification or substrate availability~\cite{mischkulnig_heme_2022}. The absence of a statistically significant negative correlation in the unadjusted proteomic analysis of CPTAC-GBM ($n=99$) reinforces that correlations seen at the mRNA level do not necessarily translate at the protein level: the negative correlation between MAPK activity and \textbf{PPOX} was also found at the protein level but did not reach statistical significance ($\rho = -0.13$, 95\% CI $[-0.32, 0.07]$; $p = 0.21$). Attenuation could be due to the smaller sample size of the CPTAC-GBM dataset ($n = 99$ compared to $133/87$), lower sensitivity of mass-spectrometry-based proteomics with respect to small effect sizes, and post-translational dissociation between protein and transcript abundance levels. In line with this, a previous study found no difference in mRNA levels between fluorescent and non-fluorescent gliomas, but found a significantly higher level of PPOX protein in fluorescent tumors~\cite{mischkulnig_heme_2022}, suggesting the relevance of PPOX to fluorescence may be better reflected at the protein level. Therefore, the correlations mentioned above cannot establish causality. Survival and subtype data were only available for TCGA, with two subtypes having few samples (neural, $n=6$; proneural, $n=12$). PPOX was not manipulated and PpIX production was not directly measured. Hence, the association between PPOX expression and MAPK signaling in glioblastoma is best viewed as hypothesis generating rather than confirmatory.

\section{Conclusion}
\label{sec:conclusion}
In an attempt to investigate the presence of the previously described ABCB1/FECH-mediated MAPK mechanism for PpIX processing in human glioblastoma~\cite{chelakkot_mek_2020}, we used a discovery-validation model in two independent, purity-adjusted cohorts. Neither cohort showed an association between ABCB1 and MAPK, while only one cohort showed an association between FECH and MAPK. The more global approach to all the enzymes involved in heme biosynthesis revealed that PPOX had a significant negative correlation with MAPK activation in both cohorts, advancing the candidate mechanism from PpIX removal toward its synthesis while supporting the hypothesis that increased MAPK activity lowers tumor fluorescence. As this is a correlation study, we will have to: (1) manipulate MAPK pathway activation and PPOX expression levels (knockdown or overexpression), and (2) correlate this with PpIX levels and 5-ALA-induced fluorescence. Note that the translational lever that this model actually implies is not direct augmentation of PPOX, but MEK/MAPK inhibition (e.g., trametinib or selumetinib). This is the first replicated tissue-level evidence linking MAPK pathway activity to the synthetic step of PpIX formation (PPOX) rather than to clearance, shifting the question from whether the link exists to how it might be exploited in the surgical setting.

\section*{Acknowledgments}
The results published here are in whole or part based upon data generated by the TCGA Research Network (\url{https://www.cancer.gov/tcga}), the Chinese Glioma Genome Atlas, and the Clinical Proteomic Tumor Analysis Consortium.

\section*{Code and Data Availability}
Code and complete robustness results, including a detailed accounting of sample exclusions for both cohorts, are available from the author upon reasonable request.

\section*{Ethics Statement}
This study used publicly available, de-identified data from TCGA, CGGA, and CPTAC. No new human or animal subjects research was conducted, and institutional review board approval was therefore not required.

\section*{Competing Interests}
The author declares no competing interests.

\section*{Funding}
This research received no specific grant from any funding agency in the public, commercial, or not-for-profit sectors.

\section*{Author Contributions and Responsibility}
All study design, data analysis, interpretation, and scientific claims are the author's own, and the author takes full responsibility for the content of this manuscript.

{\small
\bibliographystyle{unsrtnat}
\bibliography{references}
}
\end{document}